\documentclass{ifacconf}

\usepackage{graphicx}      
\usepackage{natbib}        
\usepackage[utf8]{inputenc}
\usepackage[T1]{fontenc}
\usepackage{eucal}
\usepackage{microtype}
\usepackage{csquotes}
\expandafter\let\csname equation*\endcsname\relax
\expandafter\let\csname endequation*\endcsname\relax
\usepackage{mathtools}
\usepackage{amssymb,amsfonts,stmaryrd,mathtools} 
\usepackage{mathrsfs}  
\usepackage{tikz}
\usetikzlibrary{cd}
\usepackage{color,soul}
\usepackage{enumitem}
\usepackage[normalem]{ulem}
\setlist[enumerate,1]{label={\roman*)}}
\newcommand*{\dd}{\mathrm{d}}

\newcommand{\X}{\mathfrak{X}} 
\newcommand*{\contr}[1]{\iota_{#1}}

\newcommand{\RR}{\mathbb{R}}
\newcommand{\Sp}{\mathbb{S}}

\newcommand{\Cinfty}{\mathscr{C}^{\infty}}
\newcommand{\T}{\mathrm{T}}
\newcommand{\cT}{\mathrm{T}^{\ast}}

\DeclareMathOperator{\sn}{sn}
\DeclareMathOperator{\cn}{cn}

\DeclareMathOperator{\Ima}{Im}

\newcommand\restr[2]{{
  \left.\kern-\nulldelimiterspace 
  #1 
  \right|_{#2} 
}}

\newtheorem{theorem}{Theorem}
\newtheorem{definition}{Definition}

\begin{document}
\begin{frontmatter}

\title{On the integrability of hybrid\\ Hamiltonian systems\thanksref{footnoteinfo}} 

\thanks[footnoteinfo]{The authors acknowledge financial support from the Spanish Ministry of Science and Innovation (MCIN/AEI/ 10.13039/501100011033), under grants PID2022-137909NB-C21 and RED2022-134301-T. A.~L.-G.~also received finance from the grant CEX2019-000904-S and the predoctoral contract PRE2020-093814.
}

\author[First]{Asier López-Gordón} 
\author[Second]{Leonardo J.~Colombo}

\address[First]{Institute of Mathematical Sciences, Spanish National Research Council (CSIC), Madrid, Spain (e-mail: asier.lopez@icmat.es).}
\address[Second]{Centre for Automation and Robotics, Spanish National Research Council (CSIC), Arganda del Rey, Madrid, Spain (e-mail: leonardo.colombo@car.upm-csic.es)}

\begin{abstract}                
A hybrid system is a system whose dynamics are controlled by a mixture of both continuous and discrete transitions. 
The integrability of Hamiltonian systems is often identified with complete integrability or Liouville integrability, that is, the existence of as many independent integrals of motion in involution as the dimension of the phase space. Under certain regularity conditions, Liouville--Arnold theorem states that the invariant geometric structure associated with Liouville integrability is a fibration by Lagrangian tori, on which motions are linear. In this paper, we study an extension of the Liouville--Arnold theorem for hybrid systems whose continuous dynamics is given by a Hamiltonian vector field and we state conditions on the impact map and the switching surface under which a hybrid Hamiltonian system is, in a certain sense, completely integrable.
\end{abstract}

\begin{keyword}
Hybrid systems, Integrability, Arnold-Liouville theorem, action-angle coordinates.
\end{keyword}

\end{frontmatter}


\section{Introduction}
A large class of physical systems can be described through Hamiltonian systems (see for instance \cite{Abraham2008,Arnold1978,Libermann1987,Wiggins2003}). In such systems, the total energy is described by a Hamiltonian, a smooth scalar function of the system’s generalized coordinates. The dynamics is then completely determined by the partial derivatives of the Hamiltonian with respect to the generalized coordinates. These equations are called Hamilton equations, a system of first-order differential equations. A Hamiltonian system is integrable if there exist explicit solutions to Hamilton’s equations of motion. Liouville–Arnold theorem (see \cite{Arnold1978,Bolsinov2004,Wiggins2003}) asserts that provided that there exist some functionally independent functions called first integrals, which are in involution under a Poisson bracket, the equations of motion of a Hamiltonian system can be integrated by quadratures. That is, Hamilton equations can be solved by a finite number of algebraic operations and the calculation of integrals of known functions. More formally, given a symplectic manifold of dimension $2n$, a Hamiltonian system is (Liouville) completely integrable if one can find n independent functions that pairwise Poisson-commute and are independent on a dense open subset. In this situation, this open subset admits a Lagrangian foliation and the solutions of the dynamics live in the leaves of the foliation. This notion may be extended, in a natural way, for the more general case when the phase space is a Poisson or Dirac manifold, not necessarily symplectic.

Hybrid systems are dynamical systems with continuous-time and discrete-time components on its dynamics. This class of dynamical systems  are capable of modelling several physical systems, such as multiple UAV (unmanned aerial
vehicles) systems (see \cite{lee_geometric_2013}) and legged robots  (see \cite{westervelt_feedback_2018}), among many others (see for instance \cite{goebel_rafal_hybrid_2012}, \cite{schaft_introduction_2000}). A simple hybrid system is characterized by a tuple $\mathcal{H}=(M, X, S, \Delta)$, where $M$ is a smooth manifold, $X$ is a smooth vector field on $M$, $S$ is an embedded submanifold of $M$, and $\Delta:S\to D$ is a smooth embedding. This type of hybrid system was introduced in \cite{johnson_simple_1994}, and it  has been mainly employed for the understanding of locomotion gaits in bipeds and insects (see for instance \cite{ames_geometric_2007}, \cite{holmes_dynamics_2006}, \cite{westervelt_feedback_2018}). In the situation where the vector field $X$ is associated with a mechanical system (Lagrangian or Hamiltonian), alternative approaches for mechanical systems with nonholonomic and unilateral constraints have been considered in \cite{clark_bouncing_2019}, \cite{cortes_mechanical_2001}, \cite{cortes_hamiltonian_2006},
\cite{ibort_mechanical_1997},
\cite{ibort_geometric_2001}, \cite{colombo2020symmetries},  \cite{colombo2022contact}. 

This paper aims to provide a first approach to a formal definition of complete integrability for hybrid Hamiltonian systems and a way to construct action-angle variables for these systems. To do that, we will make use of the so-called generalized hybrid momentum map and hybrid constants of the motion introduced in our previous works (see \cite{colombo2021generalized}).

The structure of the paper is as follows. In Section \ref{sec2}, we review Hamiltonian systems and complete integrable Hamiltonian systems. Section \ref{sec3} introduces hybrid Hamiltonian systems, hybrid constants of the motion and generalized hybrid momentum maps. After that, we state the Liouville--Arnold theorem for hybrid Hamiltonian systems. Finally, Section \ref{sec5} presents two examples of application: a rolling disk hitting a wall and a pendulum hitting a surface.

\textbf{Notation and conventions.} Henceforth, unless otherwise stated, all structures to be considered are assumed to be smooth. Manifolds are assumed to be Hausdorff and second-countable. All neighbourhoods are assumed to be open. Sum over crossed repeated indices is understood. 

Given a manifold $M$, $\mathfrak{X}(M)$ and $\Omega^k(M)$ will denote the space of vector fields and the space of differential $k$-forms on $M$, respectively. For $X\in\X(M)$ and $\alpha\in \Omega^k(M)$,
the exterior derivative of $\alpha$ is denoted by $\dd \alpha$, and the contraction of $\alpha$ with $X$ is denoted by $\contr{X}\alpha$. 
The tangent and cotangent bundles of a manifold $M$ will be denoted by $\T M$ and $\cT M$, respectively. Given two manifolds $M$ and $N$, and a map $F\colon M \to N$, the tangent map of $F$ will be denoted by $\T F \colon \T M \to \T N$.

\section{Integrability of Hamiltonian systems}\label{sec2}
\subsection{Hamiltonian systems}
Let $Q$ be an $n$-dimensional manifold, representing the space of positions of a mechanical system. Let $\cT Q$ denote its cotangent bundle, with canonical projection $\pi_Q: \cT Q\to Q$. If $Q$ has local coordinates $(q^i)$, then $\cT Q$ has induced bundle coordinates $(q^i, p_i)$. As it is well-known, the cotangent bundle is endowed with a canonical one-form $\theta_Q= p_i \dd q^i$, which defines a canonical symplectic form $\omega_Q=-\dd \theta_Q=\dd q^i \wedge \dd p_i$. 

For each function $H$ on $\cT Q$, the symplectic form $\omega_Q$ defines a unique vector field $X_H$ on $\cT Q$ given by    $\omega_Q\left(X_H, \cdot \right) = \dd H$, which is called the \emph{Hamiltonian vector field} of $H$. Locally,
\begin{equation}
    X_H = \frac{\partial H}{\partial p_i} \frac{\partial}{\partial q^i} - \frac{\partial H}{\partial q^i} \frac{\partial}{\partial p_i}\, , 
\end{equation}
whose integral curves are given by \textit{Hamilton's equations}
\begin{equation}
\frac{\dd q^i}{\dd t} = \frac{\partial H}{\partial p_i},\quad \frac{\dd p_i}{\dd t} = - \frac{\partial H}{\partial q^i}\, .
\end{equation}
Therefore, the dynamics of a mechanical system on $Q$ with Hamiltonian function $H: \cT Q \to \mathbb{R}$ can be characterized by the Hamiltonian vector field $X_H$.
The triple $(\cT Q, \omega_Q, H)$ is called a \emph{Hamiltonian system}. In a more general setting, one may consider a symplectic manifold which is not the cotangent bundle with the canonical symplectic form.

\subsection{Completely integrable Hamiltonian systems}
Roughly speaking, a completely integrable system is a mechanical system with $n$ independent and ``compatible'' constants of the motion, where $n$ is the number of degrees of freedom of the system.  In such systems, the equations of motion can be completely ``solved'', being reduced to quadratures.

Recall that the Poisson bracket 
is given by
$$\{f, g\} = \omega_{Q}(X_f, X_g)\, ,$$
for each pair of functions $f, g\in \Cinfty(\cT Q)$, where $X_f$ and $X_g$ denote their corresponding Hamiltonian vector fields.
A collection of functions $f_1, \ldots, f_n\in \Cinfty(\cT Q)$ are said to be \emph{in involution} if $\{f_i, f_j\}=0$ for each $i, j=1,\ldots, n$. For a Hamiltonian system $(\cT Q, \omega_Q, H)$, a function $f\in \Cinfty(\cT Q)$ is a conserved quantity if and only if it is in involution with $H$.

 A submanifold $N\subset \cT Q$ is called \textit{Lagrangian} if $\dim N = n$ and $\restr{\omega_Q}{N}=0$.

  A Hamiltonian system is called \textit{completely integrable} (or \textit{Liouville integrable}) if there exists $n$ functions $f_1,\, f_2,\, \ldots,\, f_n\in \Cinfty(\cT Q)$ such that
        \begin{enumerate}
            \item $H,\, f_1,\, f_2,\, \ldots,\, f_n$ are in involution,
            \item they are functionally independent (i.e.~$\dd f_1 \wedge \cdots \wedge \dd f_n\neq 0$) almost everywhere,
        \end{enumerate}
        The functions $f_1,\, f_2,\, \ldots,\, f_n$ are called \textit{integrals}.

Consider a completely integrable Hamiltonian system. Let $M_\Lambda$ be a regular level set of the integrals $f_1, \ldots, f_n$, i.e.
        $$M_\Lambda = \cap_{i=1}^n f_i^{-1} (\Lambda_i)
        \, ,\quad 
        \dd_x f_1 \wedge \cdots \wedge \dd_x f_n \neq 0\ \forall\, x \in M_\Lambda\, .$$
        Then, Liouville--Arnold theorem (see \cite{Arnold1978} or \cite{Wiggins2003} for instance) states that 
        \begin{enumerate}
            \item Each regular level set $M_\Lambda$ is a Lagrangian submanifold of $\cT Q$, and it is invariant with respect to the flows of $X_H, X_{f_1}, \ldots, X_{f_n}$.
            \item Any compact connected component of $M_\Lambda$ is diffeomorphic to an $n$-dimensional torus $\mathbb{T}^n$.
            \item On a neighbourhood of $M_\Lambda$ there are coordinates $(\varphi^i, s_i)$ such that
            \begin{enumerate}
                \item they are Darboux coordinates for $\omega$, namely, $\omega = \dd \varphi^i \wedge \dd s_i$,
                \item the action coordinates $s_i$ are functions depending only on the integrals $f_1, \ldots, f_n$,
                \item the integral curves of $X_H$ are given by
                 $$\dot \varphi^i = \Omega^i(s_1, \ldots, s_n),\qquad \dot s_i = 0\, .$$
            \end{enumerate}
        \end{enumerate}

The coordinates $(\varphi^i)$ and $(s_i)$ are called \emph{angle} and \emph{action coordinates}, respectively. The action coordinates determine the invariant submanifold $M_{\Lambda}$ in which the motion takes place; while the angle coordinates describe the motion along $M_{\Lambda}$, which has constant angular velocity depending only on the level set $M_{\Lambda}$.

\section{Integrability of Hybrid Hamiltonian Systems}\label{sec3}

\subsection{Hybrid Hamiltonian Systems}
Hybrid dynamical systems are dynamical systems characterized by their mixed behavior of continuous and discrete dynamics where the transition is determined by the time when the continuous flow switches from the ambient space to a submanifold. This class of dynamical systems is given by a $4$-tuple $(M,X, S, \Delta)$ formed by a manifold $M$, a vector field $X\in \X(M)$, a submanifold $S$ of $M$ and an embedding $\Delta\colon S\to M$. 
The pair $(M, X)$ describes the continuous dynamics as \begin{equation*}
\dot{x}(t) = X(x(t))
\end{equation*} 
while $(S,\Delta)$ describes the discrete dynamics as 
$x^{+}=\Delta(x^{-})$. 
The submanifold $S$ and the embedding $\Delta$ are called the \emph{impact surface} and the \emph{impact map}, respectively.

The hybrid dynamical system describing the combination of both dynamics is given by
\begin{equation}\label{sigma}
\Sigma:
\left\{ \begin{array}{ll}
\dot{x}(t) = X\big(x(t)\big)\, ,& x(t)\not\in S\, ,\\
x^+(t) = \Delta\big(x^-(t)\big)\, ,& x^-(t)\in S\, ,
\end{array}
\right.
\end{equation}
where $x^{-}$ and $x^{+}$ denote the states immediately before and after the times when $x(t)$ intersect with $S_H$, namely 
$$x^{-}(t)\coloneqq \displaystyle{\lim_{\tau\to t^{-}}}x(\tau), \qquad 
x^{+}(t)\coloneqq \displaystyle{\lim_{\tau\to t^{+}}}x(\tau)\, .$$
The flow of the hybrid dynamical system \eqref{sigma} is denoted by $\varphi_t^H$. This may cause a little confusion around the break points, that is, where $\varphi_{t_0}(x)\in S$. It is not clear whether $\varphi_{t_0}^H(x) = \varphi_{t_0}(x)$ or $\varphi_{t_0}^H(x) = \Delta(\varphi_{t_0}(x))$. In other words, if the state at the time of impact with the submanifold $S$ is $x^-$ or $x^+$. We will take the second value, i.e. $\varphi_{t_0}^H(x)=x^+$.

A solution of a hybrid dynamical system may experience a Zeno state if infinitely many impacts occur in a finite amount of time. To exclude these types of situations, we require the set of impact times to be closed and discrete, as in \cite{westervelt_feedback_2018}, so we will assume implicitly throughout the remainder of the paper that $\overline{\Delta}({S})\cap{S}=\emptyset$ (where $\overline{\Delta}({S})$ denotes the closure of $\Delta({S})$) and that the set of impact times is closed and discrete. 

Let $(M,X, S, \Delta)$ be a hybrid dynamical system.
A function $f$ on $M$ is called a \emph{hybrid constant of the motion} if it is preserved by the hybrid flow, namely, $f \circ \varphi_t^H = f$. In other words,
$X(f)=0$ and $ f \circ \Delta_H = f \circ i$, where $i\colon S \hookrightarrow M$ is the canonical inclusion.

A  hybrid dynamical system $(M,X,S,\Delta)$ is said to be a \emph{hybrid Hamiltonian system} if $M=\cT Q$ is the cotangent bundle of a manifold $Q$ and $X=X_H$ is the Hamiltonian vector field of a function $H\in \Cinfty(\cT Q)$ with respect to the canonical symplectic structure.


\subsection{Generalized hybrid momentum maps}

 Momentum maps capture in a geometric way conserved quantities associated with symmetries. 
 
 Let $G$ be a Lie group with Lie algebra $\mathfrak{g}$, and let $Q$ be a manifold.
Given a Lie group action $\psi \colon G \times Q \to Q$ of $G$ on $Q$, there is a natural lift $\psi^{\cT Q}\colon G \times \cT Q \to \cT Q$, the \textit{cotangent lift}, defined by $(g,(q,p))\mapsto (\cT \psi_{g^{-1}}(q,p))$. For each $\xi\in \mathfrak{g}$, there is a vector field $\xi_{\cT Q}\in \X(\cT Q)$ given by
$$\xi_{\cT Q}(x)=\frac{\dd}{\dd\epsilon}\psi^{\cT Q}\big(\exp(\epsilon\xi), x\big)\Big{|}_{\epsilon=0}\, ,$$
where $\exp\colon \mathfrak{g}\to G$ denotes the exponential map.
A map $\mathbf{J}:\cT Q\to\mathfrak{g}^\ast$ is called a \textit{momentum map} if $X_{\langle\mathbf{J},\xi\rangle}=\xi_{\cT G}$ for each $\xi\in\mathfrak{g}$, where $\langle\mathbf{J},\xi\rangle(x)=\langle\mathbf{J}(x),\xi\rangle$, and 
$X_{\langle\mathbf{J},\xi\rangle}$ denotes the Hamiltonian vector field of $\langle\mathbf{J},\xi\rangle$ with respect to the canonical symplectic structure. In other words, $\mathbf{J}\colon \cT Q\to\mathfrak{g}^\ast$ is a momentum map if and only if
\begin{equation}
    \contr{\xi_{\cT Q}} \omega_Q = \dd \big(\langle\mathbf{J},\xi\rangle\big)\, ,
\end{equation}
for each $\xi \in \mathfrak{g}$.

Noether's theorem states that if $H$ is a $G$-invariant Hamiltonian function on $\cT Q$ then $\mathbf{J}$ is conserved on trajectories of the Hamiltonian vector field $X_H$.

By a \textit{hybrid action} on the simple hybrid Hamiltonian system $\mathcal{H}_H$  we mean a Lie group  action $\psi\colon G\times Q\to Q$ such that
\begin{itemize}
	\item $H$ is invariant under $\psi^{\cT Q}$, i.e. $H\circ \psi^{\cT Q}=H$,
	\item $\psi^{\cT Q}$ restricts to an action of $G$ on $S_{H}$,
	\item $\Delta_{H}$ is equivariant with respect to the previous action, namely $$\Delta_{H}\circ \restr{\psi^{\cT Q}_g}{S_{H}}=\psi^{\cT Q}_g\circ \Delta_{H}.$$
\end{itemize}

A momentum map $\mathbf{J}$ will be called a \textit{generalized hybrid momentum map} for $\mathcal{H}_H$ if, for each regular value $\mu_-$ of $\mathbf{J}$ and each connected component $C$ of $S_H$,
\begin{equation}
  \Delta_{H} \left(\mathbf{J}|_{C}^{-1}(\mu_-)  \right) \subset \mathbf{J}^{-1}(\mu_+),
  \label{generalized_hybrid_momentum}
\end{equation}
for some regular value $\mu_+$. In other words, for every point in the connected component of the switching surface such that the momentum before the impact takes a value of $\mu_-$, the momentum will take a value $\mu_+$ after the impact. That is, the switching map  translates the dynamics from one level set of the momentum map into another.
In particular, when $\mu_+=\mu_-$ for each $\mu^-$ (i.e., $\Delta_{H}$ preserves the momentum map), $\mathbf{J}$ is called \textit{hybrid momentum map} (see \cite{ames_hybrid_2006}). 

Given an action in the Lie algebra such that it preserves the Hamiltonian function and is equivariant with respect to the impact map. The \textit{hybrid Noether theorem} states that for all $\xi\in\mathfrak{g}$, the function $\langle\mathbf{J}, {\xi}\rangle$, defined by contracting of the generalized hybrid momentum map with $\xi$, is a hybrid constant of the motion. 

\subsection{Completely integrable hybrid Hamiltonian Systems}

If a $2n$-dimensional Hamiltonian system $(\cT Q, \omega_Q, H)$ is completely integrable, then the Hamiltonian flows of the integrals $f_1, \ldots, f_n$ define an Abelian Lie group action of $\RR^n$ on $\cT Q$. In other words, the Hamiltonian vector fields $X_{f_1}, \ldots, X_{f_n}$ are the infinitesimal generators of the action. Hence, the momentum map can be identified with the map $F=(f_1, \ldots, f_n)\colon \cT Q\to \RR^n$. The combination of this with the notion of generalized hybrid momentum map motivates the following definition.

\begin{definition}
    Let $(M,S,X,\Delta)$ be a hybrid dynamical system. A function $f \colon M \to \RR$ is called a \emph{generalized hybrid constant of the motion} if 
    \begin{enumerate}
        \item $X f = 0$,
        \item For each connected component $C\subseteq S$ and each $a\in \Ima f$, there exists a $b\in \Ima f$ such that
        \begin{equation}\label{eq:condition_generalized_hybrid_constant}
            \left(\restr{f\circ \Delta}{C}\right)^{-1} (a) \subseteq f^{-1}(b)\, .
        \end{equation}
    \end{enumerate}
    In other words, the value of $f$ after an impact in $C$ is uniquely determined by its value before the impact.
\end{definition}

\begin{definition}
Let $Q$ be an $n$-dimensional manifold.
A \emph{completely integrable hybrid Hamiltonian system} is a $5$-tuple\newline
$(\cT Q, S, X_H, \Delta, F)$, formed by a hybrid Hamiltonian system $(\cT Q, S, X_H, \Delta)$, together with a function $F=(f_1, \ldots, f_n)\colon \cT Q \to \RR^n$ such that $\operatorname{rank} \T_x F = n$ almost everywhere and the functions $f_1, \ldots, f_n$ are generalized hybrid constant of the motion and in involution, i.e. $\{f_i, f_j\}=0$ for all $i, j\in \{1, \ldots, n\}$.
\end{definition}

On a neighborhood $U_\Lambda$ of each regular level set $M_\Lambda = F^{-1}(\Lambda)$ there exists action-angle coordinates $(\varphi^i, s_i)$ such that the dynamics are given by
\begin{equation}
    \dot{\varphi}^i(t) = \Omega^i(s_1, \ldots, s_n)\, ,\quad  \dot{s_i} = 0\, ,
\end{equation}
for $(\varphi^i(t), s_i)\in U_\Lambda \setminus S$. On the other hand, condition~\eqref{eq:condition_generalized_hybrid_constant} implies that, for each level set $M_\Lambda$ and each connected component $C\subseteq S$, there exists a $\Lambda'\in \RR^n$ such that $\Delta (M_\Lambda\cap C) \subset M_{\Lambda'}=F^{-1}(\Lambda')$. In other words, the impact map takes invariant level sets of $F$ into level sets. Since each level set is uniquely determined by the action coordinates $s_i$ and vice versa, this implies that the value of the action coordinates after the impact depends exclusively on their value before the impact and, possibly, the connected component of the impact map where the impact occurs, but they are independent of the value of the angle coordinates. Summarizing, we have the following.

\begin{theorem}
    Consider a completely integrable hybrid Hamiltonian system $(\cT Q, S, X_H, \Delta)$, with $F=(f_1, \ldots, f_n)$, where $n=\dim Q$. Let $M_{\Lambda}$ be a regular level set of $F$, namely, $M_\Lambda= F^{-1}(\Lambda)$ for $\Lambda\in \RR^n$ such that $\operatorname{rank} \T_x F = n$ for all $x\in M_{\Lambda}$. Then:
    \begin{enumerate}
        \item Each regular level set $M_\Lambda$ is a Lagrangian submanifold of $\cT Q$, and it is invariant with respect to the flows of $X_H, X_{f_1}, \ldots, X_{f_n}$.
        \item Any compact connected component of $M_\Lambda$ is diffeomorphic to an $n$-dimensional torus $\mathbb{T}^n$.
        \item For each regular level set $M_\Lambda$ and each connected component $C\subseteq S$, there exists a $\Lambda'\in \RR^n$ such that $\Delta(M_\Lambda \cap C) \subset M_{\Lambda'} = F^{-1}(\Lambda')$. 
        \item On a neighbourhood $U_\lambda$ of $M_\Lambda$ there are coordinates $(\varphi^i, s_i)$ such that
        \begin{enumerate}
            \item they are Darboux coordinates for $\omega$, namely, $\omega = \dd \varphi^i \wedge \dd s_i$,
            \item the action coordinates $s_i$ are functions depending only on the integrals $f_1, \ldots, f_n$,
            \item the continuous part hybrid dynamics are given by
             $$\dot \varphi^i = \Omega^i(s_1, \ldots, s_n),\qquad \dot s_i = 0\, .
             $$
             \item In these coordinates, for each connected component $C\subseteq S$, the impact map reads $\Delta \colon (\varphi^i_-, s_i^-)\in M_\Lambda \cap C\mapsto  (\varphi^i_+, s_i^+)\in M_{\Lambda'}$, where $s_1^+, \ldots, s_n^+$ are functions depending only on $s_1^-, \ldots, s_n^-$.
        \end{enumerate}
    \end{enumerate}
\end{theorem}
 
\section{Examples}\label{sec5}

\subsection{Rolling disk with a harmonic potential hitting fixed walls}

Consider a homogeneous circular disk of radius $R$ and mass $m$ moving in the plane. The configuration space is $Q = \RR^2 \times \Sp^1$, with canonical coordinates $(x, y, \theta)$. The coordinates $(x, y)$ represent then position of the center of the disk, while the coordinate $\theta$ represents the angle between a fixed reference point of the disk and the $y$-axis.
Suppose that the Hamiltonian function $H\colon \cT Q \to \RR$ of the system is
\begin{equation}\label{eq:Hamiltonian_disk_oscillator}
    H = \frac{1}{2m} (p_x^2 + p_y^2) + \frac{1}{2mk^2} p_\theta^2 + \frac{1}{2} \Omega^2 (x^2+y^2)\, ,
\end{equation}
where $(x, y, \theta, p_x, p_y, p_\theta)$ are the bundle coordinates in $\cT(\RR^2\times \Sp^1)$. 
Consider that there are two rough walls situated at $y=0$ and at $y=h>R$. Assume that the impact with a wall is such that the disk rolls without sliding and that the change of the velocity along the $y$-direction is characterized by an elastic constant $e$. Then, the switching surface is
\begin{equation}
\begin{aligned}
    S 
    & = \left\{\left(x, R, \theta, p_x, p_y, \frac{k^2}{R} p_x\right)\mid x, p_x, p_y \in \RR, \, \theta \in \Sp^1 \right\}\\
    & \cup \left\{\left(x, h- R, \theta, p_x, p_y, \frac{k^2}{R} p_x\right)\mid  x, p_x, p_y \in \RR, \, \theta \in \Sp^1 \right\}
\end{aligned}
\end{equation}
and the impact map $\Delta\colon S\to \cT Q$ is given by 
\begin{equation}\label{eq:impact_map_disk}
    \left(p_x^{-}, p_y^{-}, p_{\theta}^{-}\right) \mapsto\left(\frac{R^2 p_x^{-}+k^2 R p_{\theta}^{-}}{k^2+R^2},-e p_y^{-}, \frac{R p_x^{-}+k^2 p_{\theta}^{-}}{k^2+R^2}\right)
\end{equation}

For simplicity's sake, let us hereafter take $m=R=k=\Omega=1$. 
The functions
\begin{equation}\label{eq:action_coords_disk}
    f_1 = \frac{p_x^2 + x^2}{2}\, , \quad
    f_2 = \frac{p_y^2 + y^2}{2}\, , \quad
    f_3 = \frac{p_\theta^2}{2}\, ,
\end{equation}
are conserved quantities with respect to the Hamiltonian dynamics of $H$. Moreover, they are in involution and functionally independent almost everywhere. 

Let $F=(f_1, f_2, f_3)\colon \cT(\RR^2 \times \Sp)\to \RR^3$. It is clear that, for $\Lambda\neq 0$, the level sets $F^{-1}(\Lambda)$ are diffeomorphic to $\Sp \times \Sp \times \RR$.
In the intersection of their domains of definition, the functions
\begin{equation}\label{eq:angle_coords_disk}
    \phi^1 = \arctan\left(\frac{x}{p_x}\right)\, , \quad 
    \phi^2 = \arctan\left(\frac{y}{p_y}\right)\, , \quad 
    \phi^3 = \frac{\theta}{p_\theta}
\end{equation}
are coordinates on each level set $F^{-1}(\Lambda)$ for $\Lambda\neq 0$. Moreover, observe that $(\phi^i, f_i),\, i\in \{1, 2, 3\}$ are Darboux coordinates for the canonical symplectic form $\omega_Q$, namely, $\omega_Q = \dd \phi^i\wedge \dd f_i$.
In these coordinates, the Hamiltonian function reads
\begin{equation}
    H = f_1 + f_2 + f_3\, .
\end{equation}
Hence, its Hamiltonian vector field is simply
\begin{equation}
    X_H = \frac{\partial}{\partial\phi^1} + \frac{\partial}{\partial\phi^2} + \frac{\partial}{\partial\phi^3}\, .
\end{equation}


From equations~\eqref{eq:impact_map_disk} and \eqref{eq:action_coords_disk}, we have that
\begin{equation}
\begin{aligned}
    & f_1 \circ \Delta = \frac{1}{2}\left(\frac{R^2 p_x+k^2 R p_{\theta}}{k^2+R^2}\right)^2+\frac{x^2}{2}= \frac{p_x^2+x^2}{2}\, , \\
    & f_2 \circ \Delta =  \frac{e^2p_y^2+y^2}{2}\, , \\
    & f_3 \circ \Delta = \frac{1}{2}\left(\frac{R p_x+k^2 p_{\theta}}{k^2+R^2}\right)^2 = \frac{p_\theta^2}{2}\, ,
\end{aligned}
\end{equation}
where we have taken into account that $p_\theta = k^2 p_x/R$ for points in the impact surface $S$.
Thus, 
\begin{equation}
\begin{aligned}
    f_1 \circ \Delta = f_1\, , \quad 
    f_3 \circ \Delta = f_3\, ,
\end{aligned}
\end{equation}
that is, $f_1$ and $f_3$ are hybrid constants of the motion. On the other hand,
\begin{equation}
    \restr{f_2}{S} = \frac{p_y^2+a^2}{2}\, ,
\end{equation}
where $a=R$ or $a=h-R$, dopending on the connected component of $S$. Thus, along $S$, one can write $p_y^2 = 2 f_2 - a^2$. Hence,
\begin{equation}
    f_2 \circ \Delta = \frac{2e^2 f_2 -e^2 a^2 + a^2}{2} = e^2 f_2 + \frac{1-e^2}{2} a^2\, .
\end{equation}
Since its value after the impact only depends on its value before the impact, $f_2$ is a generalized hybrid constant on the motion. In particular, for a purely elastic impact ($e=1$), $f_2$ becomes a hybrid constant on the motion, namely, $f_2\circ \Delta = f_2$.

Let us now compute how the impact map modifies the angle coordinates. The combination of equations~\eqref{eq:impact_map_disk} and \eqref{eq:angle_coords_disk} yields
\begin{equation}
\begin{aligned}
    & \phi^1 \circ \Delta  = \arctan\left(x\frac{k^2+R^2}{R^2 p_x+k^2 R p_{\theta}}\right) = \arctan\left(\frac{x}{p_x}\right)= \phi^1 \, , \\
    & \phi^2 \circ \Delta = \arctan\left(\frac{y}{-ep_y}\right)
    = -\arctan\left(\frac{y}{ep_y}\right)\, , \\
    & \phi^3 \circ \Delta = \theta \frac{k^2+R^2}{R p_x+k^2 p_{\theta}} 
    = \frac{\theta}{p_\theta} = \phi^3 \, , 
\end{aligned}
\end{equation}
where we have once again used that $p_\theta = k^2 p_x/R$ for points in the impact surface $S$.
We can write 
\begin{equation}
    \frac{y}{p_y} = \tan \phi^2 \, ,
\end{equation}
and therefore
\begin{equation}
    \phi^2 \circ \Delta = -\arctan\left(\frac{\tan \phi^2 }{e}\right)\, .
\end{equation}

Finally, we have to write the impact surface in terms of the action-angle coordinates $(\phi^i, f_i)$. One can write $y= \sqrt{2 f_2} \sin \phi^2$.
Taking into account that $y=a$ (where $a=R$ of $a=h-R$) on $S$, we have
\begin{equation}
    \restr{\left(2f_2 \sin^2 \phi^2\right)}{S} =  a^2\, .
\end{equation}
Similarly, one can write $p_x = \sqrt{2 f_1} \cos \phi^1$, and thus
\begin{equation}
    \restr{f_3}{S} = \frac{2k^4 f_1 \cos^2 \phi^1}{R^2}\, ,
\end{equation}
using that $p_\theta = k^2 p_x/R$ on $S$. We conclude that, in the action-angle coordinates, the impact surface reads
\begin{equation}
\begin{aligned}
    S & = \left\{\left(\phi^i, f_i\right)\mid 2f_2 \sin^2 \phi^2 = R^2 \text{ and } f_3 = \frac{2k^4 f_1 \cos^2 \phi^1}{R^2} \right\} \\
    & \cup \left\{2f_2 \sin^2 \phi^2 = (h-R)^2 \text{ and } f_3 = \frac{2k^4 f_1 \cos^2 \phi^1}{R^2} \right\}\, .
\end{aligned}
\end{equation} 
The relations between the coordinates before, $(\phi^i_-, f_i^-)$, and after, $(\phi^i_+, f_i^+)$, are
\begin{equation}
\begin{array}{lll}
    \phi^1_+ = \phi^1_-\, , \quad 
    & \phi^2_+ = -\arctan\left(\frac{\tan \phi^2_-}{e}\right)\, , \quad
    & \phi^3_+ = \phi^3_-\, , \\ \\
    f_1^+ = f_1^- \, , \quad
    & f_2^+ = e^2 f_2 + \frac{1-e^2}{2} a^2\, , \quad
    & f_3^+ = f_3^-\, ,
\end{array}
\end{equation}
where $a=R$ or $a=h-R$ depending on the wall where the impact takes place.

\subsection{Pendulum hitting a surface} 

Consider
a pendulum mounted on the floor.
The configuration space is $Q=\mathbb{S}$ with generalized coordinate $\theta$. Coordinates on $\cT \mathbb{S}$ are denoted by $(\theta,p)$. The Hamiltonian function of the system $H:\cT \mathbb{S}\to\mathbb{R}$ is given by 
$$H(\theta,p)=\frac{p^{2}}{2ml^2}+mgl(1-\cos\theta)\, .$$ 
Henceforth, we will work in units such that $m=g=l=1$.
The vector field describing the continuous-time dynamics is the Hamiltonian vector field $X_H$ of $H$ with respect to the canonical symplectic structure, namely,
\begin{equation}
    X_H = p \frac{\partial}{\partial \theta} - \sin \theta \frac{\partial}{\partial p}\, .
\end{equation}
The switching surface is given by 
$$C=\{(\theta,p)\in \cT \mathbb{S}|\cos\theta=0 \hbox{ and }p\geq 0\}\, .$$
The impact map $\Delta:S\to \cT \mathbb{S}$ is given by 
\begin{equation}\label{eq:impact_map_pendulum_canonical}
    \Delta(\theta,p)=(\theta,-ep)\, ,
\end{equation}
where $e\in [0,1]$  denotes the coefficient of restitution. In particular, for a perfectly elastic impact $e=1$, and for a perfectly plastic
impact $e=0$. In other words, the coordinates before, $(\theta^-, p^-)$, and after, $(\theta^+, p^+)$, the impact are related by
\begin{equation}\label{eq:impact_map_pendulum_canonical_2}
    \theta^+ = \theta^-\, , \quad p^+ = -e p^-\, .
\end{equation}

 Therefore the system \begin{equation*}
    \Sigma_{\mathscr{H}}:
    \left\{\begin{array}{ll}\dot{\theta}(t)=\frac{p}{ml^2},\dot{p}=-mgl\sin\theta, & \hbox{ if } \cos\theta(t)\neq 0,\, p(t)> 0,\\ \theta^{+}=\theta^{-},\,p^{+}=-ep^{-},&\hbox{ if } \cos\theta(t)=0,\, p(t)\geq 0, 
    \end{array}\right.
  \end{equation*}
is a hybrid Hamiltonian system.

The Hamiltonian function is a conserved quantity with respect to the continuous dynamics, that is, $X_H (H) = 0$. Moreover, it is a hybrid constant of the motion (i.e., $H \circ \Delta = H$) if and only if the coefficient of restitution is $e=1$. 

Let us write the level sets of $H$ in terms of a parameter $\kappa$, which we can interpret as half of the energy. There are two types of invariant subamnifolds, the so-called \textit{libration} and \emph{rotation} cases, corresponding to the level sets $H^{-1}(2\kappa)$ for $\kappa< 1$ and $\kappa>1$, respectively.

In the following we will restrict to the libration case ($\kappa<1$).
In that case, the action coordinate is given by
\begin{equation}\label{eq:action_coordinate_libration}
\begin{aligned}
    J_\ell(\theta, p) & = \frac{8}{\pi}\left[\mathrm{E}\left(\frac{H(\theta,p)}{2}\right)
    \right.\\ &\quad \left.
    -\left(1-\frac{H(\theta,p)}{2}\right) \mathrm{K}\left(\frac{H(\theta,p)}{2}\right)\right]\, ,
\end{aligned}
\end{equation}
where $\mathrm{K}$ and $\mathrm{E}$ denote the complete elliptic integrals of the first and second kinds, respectively. 
The canonical coordinates as a function of $\kappa$ and the angle coordinate $\zeta_\ell$ are
\begin{equation}\label{eq:angle_coordinate_libration}
\begin{aligned}
    & \theta\left(\kappa, \zeta_{\ell}\right)=2 \arcsin \left[\sqrt{\kappa} \sn\left(\frac{2 \mathrm{K}(\kappa)}{\pi} \zeta_{\ell} \mid \kappa\right)\right]\, , \\
    & p\left(\kappa, \zeta_{\ell}\right)=2 \sqrt{\kappa} \cn\left(\frac{2 \mathrm{K}(\kappa)}{\pi} \zeta_{\ell} \mid \kappa\right)\, ,
\end{aligned}
\end{equation}
where $\sn$ and $\cn$ denote the Jacobi elliptic functions.

The manifold $\cT \mathbb{S} \simeq \mathbb{S} \times \mathbb{R}$ is foliated by leaves which are diffeomorphic to $\mathbb{S}$. The action coordinate $J_\ell$ determines the leave of the foliation, while the angle coordinate $\zeta_\ell$ is a coordinate on the leave. Hence, the condition of hybrid momentum map implies that if $\Delta (\zeta_\ell^-, J_\ell^-) =  (\zeta_\ell^+, J_\ell^+)$, then $\Delta (\tilde{\zeta}_\ell^-, J_\ell^-) =  (\tilde{\zeta}_\ell^+, J_\ell^+)$ for all $(\zeta_\ell^-, J_\ell^-), (\tilde{\zeta}_\ell^-, J_\ell^-)\in C$. In other words, the action coordinate in the instant after the impact, $J_\ell^+$, depends only on the action coordinate in the instant before the impact, $J_\ell^-$. This means that if the leaves of the foliation are invariant under the hybrid dynamics, i.e., if the initial conditions are in one leave, after the impact the system will remain in the same leave.

By equations \eqref{eq:impact_map_pendulum_canonical} and \eqref{eq:action_coordinate_libration}, this condition is verified in the case of a completely elastic impact ($e=1$). As a matter of fact, in that case the impact map does not modify the action coordinate, namely, $J_\ell^+ = J_\ell ^-$. Similarly, $\Delta$ does not change $\kappa$. By equations~\eqref{eq:impact_map_pendulum_canonical_2} and \eqref{eq:angle_coordinate_libration}, the relation between the angle coordinates before, $\zeta_\ell^-$, and after, $\zeta_\ell^+$, the impact is given by
\begin{equation}\label{eq:impact_map_pendulum_libration}
\begin{aligned}
    & \sn\left(\frac{2 \mathrm{K}(\kappa)}{\pi} \zeta_{\ell}^+ \mid \kappa\right)
    =\sn\left(\frac{2 \mathrm{K}(\kappa)}{\pi} \zeta_{\ell}^- \mid \kappa\right)\, , \\
    & \cn\left(\frac{2 \mathrm{K}(\kappa)}{\pi} \zeta_{\ell}^+ \mid \kappa\right) 
    = - \cn\left(\frac{2 \mathrm{K}(\kappa)}{\pi} \zeta_{\ell}^- \mid \kappa\right)\, ,
\end{aligned}
\end{equation}
where it has been taken into account that $\arcsin\colon [-1,1] \to \RR$ is an injective function.

In the low energy limit ($\kappa\ll 1$), the Jacobi elliptic functions behave like trigonometric functions, namely, $\sn(x| \kappa) = \sin x + \mathcal{O}(\kappa)$ and $\cn(x| \kappa) = \cos x + \mathcal{O}(\kappa)$. Hence, equations~\eqref{eq:impact_map_pendulum_libration} can be approximated by
\begin{equation}\label{eq:impact_map_pendulum_low_energy}
\begin{aligned}
    & \sin\left(\frac{2 \mathrm{K}(\kappa)}{\pi} \zeta_{\ell}^+ \right)
    = \sin\left(\frac{2 \mathrm{K}(\kappa)}{\pi} \zeta_{\ell}^- \right) + \mathcal{O}(\kappa)\, , \\
    & \cos\left(\frac{2 \mathrm{K}(\kappa)}{\pi} \zeta_{\ell}^+ \right) 
    = - \cos\left(\frac{2 \mathrm{K}(\kappa)}{\pi} \zeta_{\ell}^- \right) + \mathcal{O}(\kappa)\, ,
\end{aligned}
\end{equation}
whose solution is $\zeta^+ = \pi - \zeta^- + \mathcal{O}(\kappa)\, (\operatorname{mod} 2\pi)$.


\section{Conclusions}
We have introduced a notion of complete integrability for simple hybrid Hamiltonian systems and described how to induce action-angle coordinates on the impact map and switching surface. Two examples have been studied a rolling disk hitting walls and a pendulum hitting a surface. For future work, we would like to understand the relation between hybrid Hamiltonian systems and KAM theory for completely integrable hybrid Hamiltonian systems.

\bibliography{ifacconf}             
                                                   







\end{document}